\documentclass[proof]{revtex4}

\usepackage{graphicx}% Include figure files
\usepackage{dcolumn}% Align table columns on decimal point
\usepackage{bm}% bold math

\begin{document}

\title{Laser-Induced Magnetic Nanostructures with Tunable Topological Properties}

\author{M.Finazzi}
\email{marco.finazzi@fisi.polimi.it}
\affiliation{CNISM - Dipartimento di Fisica, Politecnico di Milano, Piazza Leonardo da Vinci 32, 20133 Milano, Italy}
\author{M.Savoini}
\affiliation{CNISM - Dipartimento di Fisica, Politecnico di Milano, Piazza Leonardo da Vinci 32, 20133 Milano, Italy}
\affiliation{Radboud University Nijmegen, Institute for Molecules and Materials, Heyendaalseweg 135,
6525 AJ Nijmegen, The Netherlands}
\author{A. R. Khorsand}
\affiliation{Radboud University Nijmegen, Institute for Molecules and Materials, Heyendaalseweg 135,
6525 AJ Nijmegen, The Netherlands}
\author{A. Tsukamoto}
\affiliation{College of Science and Technology, Nihon University, 7-24-1 Funabashi, Chiba, Japan}
\affiliation{PRESTO, Japan Science and Technology Agency, 4-1-8 Honcho Kawaguchi, Saitama, Japan}
\author{A. Itoh}
\affiliation{College of Science and Technology, Nihon University, 7-24-1 Funabashi, Chiba, Japan}
\author{L. Du\`{o}}
\affiliation{CNISM - Dipartimento di Fisica, Politecnico di Milano, Piazza Leonardo da Vinci 32, 20133 Milano, Italy}
\author{A. Kirilyuk}\author{Th.Rasing}
\affiliation{Radboud University Nijmegen, Institute for Molecules and Materials, Heyendaalseweg 135,
6525 AJ Nijmegen, The Netherlands}
\author{M. Ezawa}
\affiliation{Department of Applied Physics, University of Tokyo, Hongo 7-3-1, 113-8656, Japan}
\date{\today}

\begin{abstract}
We report the creation and real-space observation of magnetic structures with well-defined topological properties and a lateral size as low as about 150~nm. They are generated in a thin ferrimagnetic film by ultrashort single optical laser pulses. Thanks to their topological properties, such structures can be classified as Skyrmions of a particular type that does not require an externally applied magnetic field for stabilization. Besides Skyrmions, we are able to generate magnetic features with topological characteristics that can be tuned by changing the laser fluence. The stability of such features is accounted for by an analytical model based on the interplay between the exchange and the magnetic dipole-dipole interactions.
\end{abstract}


\maketitle

Skyrmions \cite{Skyrme} are particle-like solutions of wave equations characterized by a topological index which is conserved in time and plays the important role of a quantum number for particle states in the corresponding field theory. One of the most interesting characteristics of such topological states of matter resides in the robustness that they hold with respect to perturbations and disorder.

In magnetic materials Skyrmions emerge as soliton-like excitations that cannot be traced back to the ground ferromagnetic state by continuous deformations of the local magnetization field. So far they have been observed in materials with perpendicular magnetic anisotropy in connection with one of the following stabilization mechanisms: (i) four-spin exchange, leading to the formation of a Skyrmion lattice with each Skyrmion extending over a few lattice sites \cite{Heinze}; (ii) the Dzyaloshinskii-Moriya interaction (DMI) \cite{Dzyaloshinskii, Moriya}, which is active in noncentrosymmetric helimagnets, where Skyrmions have typical dimensions of few tens of nanometers \cite{Muehlbauer, Muenzer, Yu2010, Yu2011}; (iii) the long-range dipole-dipole interaction (DDI) \cite{Yu2012} stabilizing Skyrmions with a typical lateral size of the order of $1~\mu$m \cite{Grundy, Lin, Suzuki}. It has been recently found that the DDI can induce a larger and more complex variety of magnetic textures with respect to the ones that are observed in association with four-spin exchange and the DMI \cite{Yu2012}.

Despite the native state of thin films of magnetic materials with perpendicular anisotropy is characterized by stripe domain patterns, Skyrmions can be generated when a sufficiently high magnetic field is applied perpendicular to the film surface. In this case, the stripes undergo a transition to a hexagonal Skyrmion lattice \cite{Yu2010, Yu2011,Yu2012, Grundy, Suzuki}. Skyrmions, however, can also be produced individually, as demonstrated by a laser-induced magnetization reversal experiment on a TbFeCo thin film \cite{Ogasawara}, which still required a constant magnetic field to avoid the collapse of the domains produced upon illumination by single laser pulses. It has been proposed \cite{Ezawa2010} that such magnetic features could be stabilized in the form of small cylindrical domains thanks to the frustration among the exchange interaction, the magnetic dipole-dipole coupling, and the external magnetic field.

We report the creation of magnetic Skyrmions in a ferrimagnetic thin film generated without any external field with the help of an ultrashort laser pulse. In addition, by varying the laser fluence we are able to control the size of the Skyrmions and their topological quantum number, generating either Skyrmions or even states of a Skyrmion-antiSkyrmion pair, which we named Skyrmionium in analogy to similar bound particle/anti-particle states. In the following we demonstrate that a model including exchange coupling, single-ion magneto-crystal anisotropy and magnetic dipole-dipole interactions can indeed have metastable solutions with a nonvanishing topological index. We show that the stabilizing action of the externally applied reversed magnetic field exploited in Refs.~\onlinecite{Ogasawara} and \onlinecite{Ezawa2010} can be effectively replaced by the dipolar field from the magnetic film itself.

The sample is a 20~nm-thick amorphous thin alloy film of Tb$_{22}$Fe$_{69}$Co$_9$ which was deposited by magnetron sputtering on a Si$_3$N$_4$(5~nm)/AlTi(10~nm)/glass substrate and then capped with a 60~nm-thick layer of Si$_3$N$_4$. The thin metallic layer of AlTi acts as a heat sink. The film exhibits out-of-plane anisotropy with a coercive field of 1~T and a Faraday rotation between the two opposite magnetizations as high as $0.3^\circ$ at $\lambda$=632.8~nm.

The sample was prepared by irradiation, at room temperature, with 150~fs circularly polarized single laser pulses (spot diameter $\approx2~\mu$m, $\lambda= 800$~nm), to induce ultrafast magnetization reversal \cite{Khorsand}. Permanent sample damage occurs for laser fluences above 15~mJ/cm$^2$. At variance with Ref.~\onlinecite{Ogasawara} and the works reporting DMI-stabilized Skyrmions \cite{Muehlbauer, Muenzer, Yu2010, Yu2011}, no external magnetic field was applied during this step. Finally, we mapped the out-of-plane component of the local magnetization with sub-diffraction resolution by measuring \cite{Savoini2009} the Faraday rotation of linearly polarized light ($\lambda= 635$~nm) transmitted through the film from a polarization-conserving hollow pyramid tip (aperture diameter $\approx100$~nm) \cite{Biagioni} that was scanned over the sample.

\begin{figure}
\includegraphics[width=0.38\textwidth]{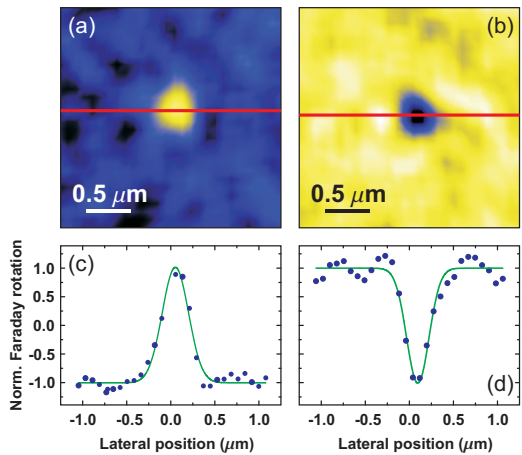}
\caption{\label{fig:1} (Color online) (a, b) Near-field Faraday rotation map showing magnetic domains induced in a thin TbFeCo film after single laser pulse irradiation (energy density $\approx5$~mJ/cm$^2)$ in film areas showing opposite out-of plane magnetizations. (c, d) Corresponding Skyrmion spin textures acccording to Eq.~(\ref{eq:4}) (with $\theta_ {0}=\pi/2$). (e, f)  Profile (dots) of the Faraday rotation measured along the lines in (b) and (c), respectively. The lines in (e) and (f) are fits performed with the trial function describing the Skyrmions discussed in the text [$R_1=200$~nm for the structure in (a) and $R_1=235$~nm for the structure in (b)].}
\end{figure}

As shown in Fig.~\ref{fig:1}, magnetic domains are obtained after single laser pulse irradiation with a pulse energy density of about 5~mJ/cm$^2$. Within experimental accuracy, these sub-$\mu$m magnetic domains display a local magnetization in their center that is completely reversed with respect to the surrounding material. Their lifetime is very long as they have not shown any noticeable change more than 12 months after the first measurements.

The mechanism stabilizing these structures resides in the interplay among the exchange coupling $H_\mathrm{X}$, the unidirectional single-ion anisotropy $H_\mathrm{A}$, and the magnetic dipole-dipole interaction $H_\mathrm{D}$ contributions to the total Hamiltonian $H=H_\mathrm{X}+H_\mathrm{A}+H_\mathrm{D}+H_\mathrm{Z}$, with $H_\mathrm{Z}$ being the Zeeman energy (see the Supplementary Material). The value of the uniaxial anisotropy term $H_\mathrm{A}$ needs to be chosen in order to also account for the presence of domain wall pinning centers, which are responsible for the high coercive fields characterizing relatively hard magnetic materials such as TbFeCo. Both magneto-crystal anisotropy and pinning centers, in fact, have a similar effect in terms of creating an energy barrier for magnetization reversal. This approach consisting in treating pinning centers as a source of anisotropy is valid as long as i) pinning sites are uniformly distributed and ii) a single Skyrmion comprises many of them. It is reasonable to assume that these two conditions apply to our case since the TbFeCo film is amorphous and the solitons that we observe have dimensions of the order of hundreds of nanometers.

In order to evaluate the existence of stable solutions to the integro-differential equations that stem from the Hamiltonian $H$, suppose there exists a soliton solution $\mathbf{n}^0(\mathbf{r})$. We can calculate the contributions to the total energy associated with $\mathbf{n}^0(\mathbf{r})$ and denote them as $E^0_\mathrm{X}$ (exchange), $E^0_\mathrm{A}$ (anisotropy), $E^0_\mathrm{D}$ (DDI) and $E^0_\mathrm{Z}$ (Zeeman). Now consider a new configuration, $\mathbf{n}(\mathbf{r})=\mathbf{n}^0(\eta \mathbf{r})$. By evaluating the scaling properties of $H_\mathrm{X}$, $H_\mathrm{A}$, $H_\mathrm{D}$, and $H_\mathrm{Z}$, we find the energy of the new configuration to be
\begin{equation}
\label{eq:1}
E_{\mathrm{tot}}(\eta) = E^0_\mathrm{X}+\eta^{-2}E^0_\mathrm{A}-\eta^{-1}|E^0_\mathrm{D}|+\eta^{-2}E^0_\mathrm{Z}.
\end{equation}
This function has a unique minimum at
\begin{equation}
\label{eq:2}
\eta = 2\frac{E^0_\mathrm{A}+E^0_\mathrm{Z}}{|E^0_\mathrm{D}|}>0.
\end{equation}
Therefore the problem is not scale-invariant and, according to the Derrick-Hobart theorem \cite{Derrick, Hobart}, may have a unique soliton solution characterized by $\eta=1$ or $2\left(E^0_\mathrm{A}+E^0_\mathrm{Z}\right)=|E^0_\mathrm{D}|$. It is apparent that the presence of perpendicular anisotropy can stabilize a soliton  even without the Zeeman term, as already suggested by Clarke et al. \cite{Clarke}. Although the role of the DDI has been questioned \cite{Kiselev}, Eq.~(\ref{eq:2}) demonstrates that the DDI is an essential requisite for the stabilization of the soliton.

We proceed by examining cylindrically symmetric general solutions of the nonlinear sigma model derived from the expression of the Hamiltonian $H$ having $n_z = \sigma(r)$, with $r$ being the distance from the soliton center and $\sigma(r)$ a trial function, $-1\leq\sigma(r)\leq+1$. The Pontryagin number $Q$ \cite{Belavin} of this spin structure is readily calculated to be $Q=[\sigma(\infty)-\sigma(0)]/2=+1$, while the total energy is finite and positive. These two conditions together with the Derrick-Hobart theorem allow us to assert the existence of a function that minimizes the total energy, even without knowing its explicit form. The winding of the spins in the structure is described by a zero-energy mode parameter $\theta_0$ (see Supplementary Material), which affects the total energy only through small corrections to the DDI term and has a negligible influence on the soliton size.

Since it is not possible to find analytical solutions, we just consider a set of reasonable trial functions and determine the one that minimizes the total energy. The line profiles in Fig.~\ref{fig:1} reporting the spatial variation of the Faraday rotation indicate that $\sigma(r)$ can be chosen as a Gaussian function,
\begin{equation}
\label{eq:4}
\sigma(r) = 1- 2e^{-r^2/R_1^2},
\end{equation}
resulting in a spin texture with $Q = 1$ (in the following we will adopt the notation $R_Q$ to indicate the size parameter for solitons with Pontryagin number equal to $Q$).

The parameters $R_1$ and $\theta_0$ can be determined by minimizing the total energy $H_\mathrm{X}+H_\mathrm{A}+H_\mathrm{D}+H_\mathrm{Z}$ of the soliton (see the Supplementary Material), giving
\begin{equation}
\label{eq:5}
R_{1} = \frac{1.62\Omega}{\pi\Gamma/\xi^2-2\Omega/d_f+2\pi\Delta_\mathrm{Z}/a^2},
\end{equation}
$d_f$ being the thickness of the film, $\Gamma$ the exchange constant, $\xi$ the single-ion uniaxial anisotropy, $\Omega$ the DDI strength, $a$ the lattice constant, and $\Delta_\mathrm{Z}$ the Zeeman energy (see the Supplementary Material). The ground state has spins forming a Bloch-type magnetic domain wall ($\theta_0=\pm\pi/2$) separating the soliton core from the surrounding medium. This soliton has a finite radius and is stable even without an external field ($\Delta_\mathrm{Z}=0$) in a sample characterized by finite uniaxial anisotropy ($\xi\neq\infty$). Our analysis of the stability of spin arrangements consistent with Fig.~\ref{fig:1} and their nonvanishing Pontryagin number allow us to consider them as genuine Skyrmions.

Another feature differentiating such Skyrmions from the disc-type ones discussed in Ref.~\onlinecite{Ezawa2010} is that the latter are stable only above a minimum size, while the size of the former can be shrunk without limit by increasing the magnetic field. By using typical sample parameters \cite{Ezawa2010,Rahman}, namely $\Gamma=3.5\times10^{20}$~J, $\xi=25$~nm, and $a=0.3$~nm, we reproduce the experimental value of the Skyrmion size in zero field, which is about 200~nm (see Fig.~\ref{fig:1}) for $\Omega=1.6\times10^{12}$~J/m, which would correspond to a saturation magnetization of about 56~emu/cm$^3$.

\begin{figure}
\includegraphics[width=0.38\textwidth]{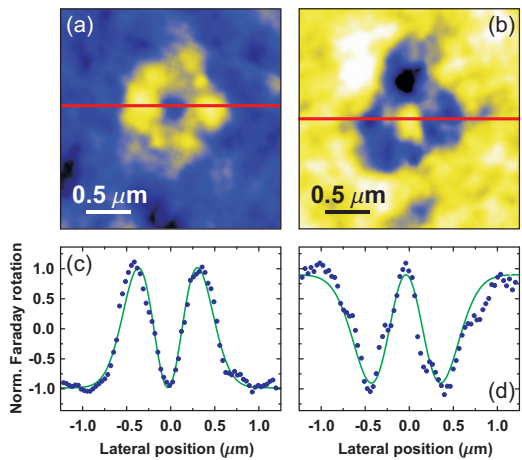}
\caption{\label{fig:3} (Color online) (a, b) Near-field Faraday rotation maps showing ``doughnut''-shaped magnetic domains after single laser pulse irradiation (energy density $\approx7$~mJ/cm$^2)$ in film areas showing opposite out-of plane magnetizations. (c, d) Corresponding Skyrmionium spin textures acccording to Eq.~(\ref{eq:6}) (with $\theta_ {0}=\pi/2$ for both Skyrmion and antiSkyrmion). (e, f) Profile (dots) of the Faraday rotation measured along the lines in (a) and (b), respectively. The full lines in (e) and (f) are fits performed with the trial function describing Skyrmionium discussed in the text [$R_0=370$~nm for the structure in (a) and $R_0=430$~nm for the structure in (b)].}
\end{figure}

At higher laser fluences with respect to those leading to Skyrmion generation (corresponding to an energy density up to 7~mJ/cm$^2$ for each single laser pulse) we observed the formation of ``doughnut''-shaped magnetic structures such as the ones shown in Figs.~\ref{fig:3}a and \ref{fig:3}b. These structures are similar to those observed after cycling the magnetization of Co/Pt multilayers with perpendicular anisotropy \cite{Iunin}: the magnetization in the center is the same as in the surrounding medium, and is completely reversed in the annular region around the core. This texture can be viewed as two Skyrmions with opposite topological numbers (i.e. a Skyrmion and an antiSkyrmion) nested into each other. This results in a state with $Q=0$, corresponding to a nontopological soliton, which can be named Skyrmionium (Sk) in analogy with states of matter constituted by a particle bound to its antiparticle. Following this notation, a Skyrmion and an antiSkyrmion can be respectively abbreviated in Sk$^+$ and Sk$^-$, since they can be obtained by ``ionizing'' Sk.

Sk is stabilized by the attractive force existing between a Sk$^+$ and a Sk$^-$, generated by their opposite magnetic dipole moments. The simplest Sk spin texture with cylindrical symmetry is given by Eq.~(\ref{eq:2}) with
\begin{equation}
\label{eq:6}
\sigma(r) = 1- 2e\left(r/R_0\right)^2e^{-r^2/R_0^2}.
\end{equation}
By substituting this trial function into our model Hamiltonian we are able to analytically estimate the value of the total energy $H_\mathrm{X}+H_\mathrm{A}+H_\mathrm{D}+H_\mathrm{Z}$ and to determine the Sk size $R_0$, which is found to be
\begin{equation}
\label{eq:7}
R_{0} = \frac{3.88\Omega}{(4-e)(\pi\Gamma/\xi^2-2\Omega/d_f)+4\pi\Delta_\mathrm{Z}/ea^2}.
\end{equation}

It is remarkable that, with no magnetic field ($\Delta_\mathrm{Z}=0$), the ratio between the Sk radius $R_0$ and the Sk$^+$ radius $R_1$ is a sample-independent quantity: $R_0/R_1=1.87$. This ratio is in excellent agreement with the experimental value obtained from Figs.~\ref{fig:1} and \ref{fig:3}: $R_0/R_1=1.8\pm0.3$. According to Eqs.~(\ref{eq:5}) and (\ref{eq:7}), both $R_1$ and $R_0$ decrease with $\Omega$ (the DDI strength), suggesting that a possible means to shrink the lateral size of such structures without the help of a magnetic field could be to employ such ferrimagnetic films at a temperature close to their magnetization compensation point, at which the two sublattices have opposite moments and the DDI vanishes.

\begin{figure}
\includegraphics[width=0.43\textwidth]{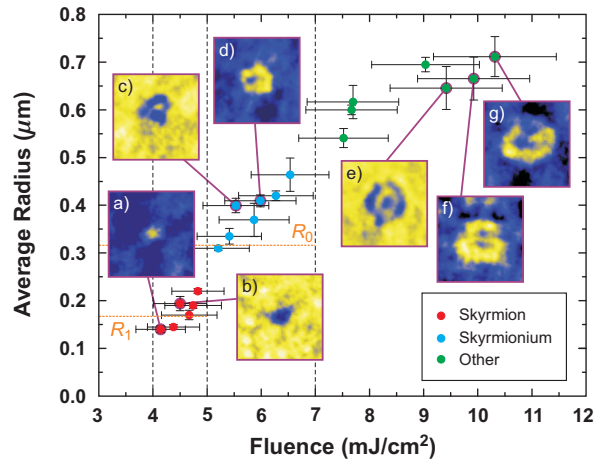}
\caption{\label{fig:4} (Color online)
Evolution of size and topology of magnetic domains generated in TbFeCo after single laser pulse illumination, as a function of laser fluence. The horizontal lines indicate the values of the $R_1$ and $R_0$ parameters evaluated for this TbFeCo sample (see text). The size is estimated by considering the rim of the domains defined by the condition of vanishing Faraday rotation and by averaging the radius of the domain over angle. The error bar on the size also considers the deviations from the axial symmetry of each considered structure. Insets: $2.5\times2.5~\mu$m$^2$ Faraday rotation maps.}
\end{figure}

The evolution of the spin textures that are generated by single-laser-pulse illumination with increasing intensity is illustrated in Fig.~\ref{fig:4}. For fluences below 4 mJ/cm$^2$ no magnetization reversal is induced, while single, isolated Skyrmions are generated in the 4-5~mJ/cm$^2$ fluence range with 100\% efficiency, i.e. each single laser pulse always gives rise to a Sk$^+$. Conversely, fluences between 5~mJ/cm$^2$ and 7~mJ/cm$^2$ cause systematic Sk formation. The diameter of both Sk$^+$ and Sk increases with fluence, suggesting that pinning defects might play a significant role in defining the ultimate size and shape of such structures. Apparently, Sk$^+$ (Sk) is systematically generated when the laser fluence is over the minimum threshold \cite{Khorsand} required to induce magnetic reversal over a wide enough area, with radius comparable to $R_1$ ($R_0$). The role of defects might thus be to prevent the structures from collapsing to the size defined by $R_1$ for Sk$^+$ or by $R_0$ for Sk.

Upon illumination with laser fluences above 7~mJ/cm$^2$, larger structures are generated, with topological properties that may vary from structure to structure. The texture shown in Fig.~\ref{fig:4}e closely resembles to those observed in hexaferrite \cite{Yu2012}. We interpret it as a combination of \textit{two} Sk$^+$ and \textit{one} Sk$^-$, resulting in a structure with $Q=+1$. It can be viewed as a di-Skyrmionium ``cation'' molecule Sk$^+_2$, representing the topological analogue of di-positronium molecules. According to our model, concentric ring structures, made of alternating $N$ (or $N+1$) Skyrmions and $N$ antiSkyrmions nested together, should all be stable. Their Skyrmion number is 0 or $\pm1$ depending on the number of Skyrmions being larger or equal to the number of antiSkyrmions. We find that the values of the zero-order parameter of adjacent (anti)Skyrmions should always be opposite, in agreement with the alternating helicity reversals inside Skyrmions with multiple-ring structure reported for hexaferrite \cite{Yu2012}.

Figure~\ref{fig:4}f shows a different di-Skyrmionium texture consisting of \textit{one} Sk$^+$ surrounding \textit{two} Sk$^-$, yielding $Q=-1$. In such a Sk$^-_2$ ``anion'' molecule the repulsion between the two Sk$^-$ having \textit{parallel} dipole magnetic moments is screened by the Sk$^+$ that encircles them, which has an \textit{opposite} dipole moment. It is still an open question whether Sk$^-_2$ is intrinsically stable or a nonuniform distribution of spin-pinning defects in the polycrystalline film is essential to prevent this structure from collapsing or expanding. The non-centrosymmetric shape of the structures shown in Fig.~\ref{fig:4} indeed suggests that defects are distorting the spin arrangement. Although we are not able to determine the conditions leading to the creation of either Sk$^+_2$ or Sk$^-_2$, their observation together with the reliable production of Skyrmionium represents an important step towards the study of complex topological states in condensed matter.

Our results clearly demonstrate that by illuminating a thin TbFeCo film with single ultrafast laser pulses we can locally create stable magnetic configurations characterized by well-defined topological numbers. By tuning the laser fluence we can control the topology and create various bound states of Skyrmion molecules (Skyrmion ``chemistry''). Such magnetic excitations are stable even without an external magnetic field. The ability of controlling the dimension and the number of Skyrmions or antiSkyrmions created during a single laser illumination event and the fact that such structures are topologically protected and thus intrinsically stable might have far reaching implications for the further development of non-volatile high-density magnetic recording and long term memories of the next generation.

\begin{acknowledgments}
The authors would like to thank A. V. Kimel for support and stimulating discussion. This research has received funding from Fondazione Cariplo through the project PONDER (Rif. 2009-2726), from MIUR through the PRIN project No.~2008J858Y7, from the European Union (EU) Nano Sci-European Research Associates (ERA) project FENOMENA, from Stichting voor Fundamenteel Onderzoek der Materie (FOM), De Nederlandse Organisatie voor Wetenschappelijk Onderzoek (NWO), and EC FP7 [Grants No.~NMP3-SL-2008-214469 (UltraMagnetron) and No.~214810 (FANTOMAS)]. Grants-in-Aid for Scientific Research from the Ministry of Education, Science, Sports and Culture of Japan No.~22740196 is also gratefully acknowledged.
\end{acknowledgments}

\end{document}